\documentclass[%
 reprint,
 amsmath,amssymb,
 aps,
 prl
]{revtex4-2}

\usepackage{graphicx}
\usepackage{dcolumn}
\usepackage{bm}
\usepackage[colorlinks=true,linkcolor=blue]{hyperref}

\begin{document}

\preprint{APS/123-QED}

\title{Polymerization in magnetic metamaterials}

\author{Samuel D. Slöetjes}
\author{Matías P. Grassi}
\author{Vassilios Kapaklis}
\affiliation{Department of Physics and Astronomy, Uppsala University, Box 516, SE-75120 Uppsala, Sweden}%

\begin{abstract}
We numerically study a mesoscopic system consisting of magnetic nanorings in the presence of thermal magnetization fluctuations. We find the formation of dipolar-field-mediated ``bonds" promoting the formation of annuli clusters, where the amount of bonds between two rings varies between zero and two. This system resembles the formation of polymers from artificial atoms, which in our case are the annuli and where the valency of the atom is set by the ring multipolarity. We investigate the thermodynamic properties of the resulting structures, and find a transition associated with the formation of the bonds. In addition, we find that the system has a tendency to form topological structures, with a distinct critical temperature in relation to the one for bond formation.
\end{abstract}

\maketitle

Polymerization is typically related to the process of reaction of monomer molecules towards the formation of larger chain-like or three-dimensional molecular networks, polymers. Essential for these processes is the ability of monomers to bond with other monomers and their steric effects, relating to the way atoms can spatially arrange. Extrapolating these observations into the realm of magnetism and artificial arrays of mesoscopic magnetic entities, allows for a new look and design approach for the collective magnetic order and dynamics of magnetic metamaterials \cite{Heyderman:2013gb,kaya2022soft}. Important to this shift in perspective, is introducing the concept of bonds between magnetic entities, captured by the dipole coupling between them and the associated magnetostatic charges. The formation of 1D chains in magnetic metamaterials has been observed before, for example in square arrays of circular disks in the form of antiferromagnetic lines \cite{digernes2020direct, sloetjes2017tailoring}, which can be regarded as polymers with a trivial topology. Another 1D entity in metamaterials consisting of bistable elements are Dirac strings, which connect certain high energy vertex configurations \cite{Mengotti:2010fo, Drisko:2017fb}. However, structures with topologies beyond that of an open chain remain scarcely explored. 

\begin{figure}
    \centering
    \includegraphics[width=\columnwidth]{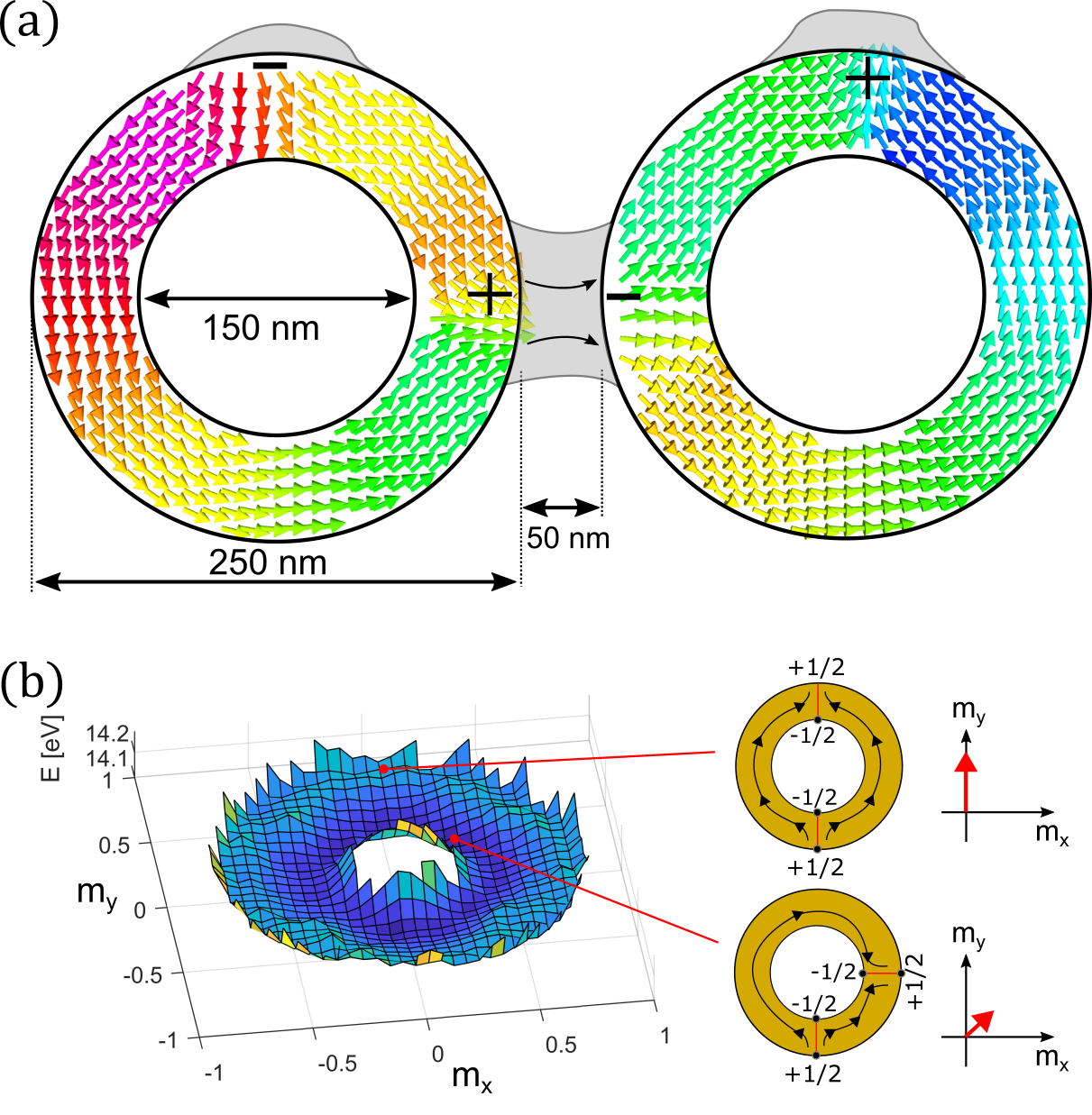}
    \caption{(a) Magnetic rings, each with two domain walls with opposite-sign magnetostatic charges \cite{Tchernyshyov:2005gs}. These charges can couple across the elements, forming bonds via stray fields (gray shaded areas). (b) Energy landscape for two domain wall states in magnetic nanorings, as a function of the net magnetic moment of the ring (indicated on the right).} 
    \label{fig:intro}
\end{figure}

The magnetic texture within the individual elements can be harnessed in order to allow for increased complexity of the emergent artificial structures in these materials. One way to achieve this, is by altering the topology of the building blocks themselves. Tailoring of the topology in magnetic metamaterials has so far only been realized on the lattice level, for the realization of frustrated magnetic systems \cite{morrison2013unhappy}, and has lead for example to the investigations of Shakti \cite{Gilbert_Shakti_2014, Stopfel:2018dw}, Saint George \cite{Stopfel_SaintGeorge}, and Tetris \cite{Gilbert_tetris_2015bp} artificial spin ice, among others. In this work, we will consider a system consisting of building blocks with an altered topology compared to the usual Ising-like (elongated) or disk-shaped mesospins (for a classification scheme see Table I by
\citet{Skovdal_PRB2021}), namely magnetic rings. Nanomagnetic rings have been studied previously, albeit in a different context. Early efforts consisted of studying domain walls in single rings \cite{rothman2001observation, klaui2003vortex, muscas2020mesoscale}, which focused on switching between micromagnetic states in the ring and subsequent dynamics investigations \cite{PhysRevB.76.014431, PhysRevB.78.104421}. Studies on the thermally driven dynamics in ring systems are scarce, and only focus on the thermally excited transition from a vortex state to an onion state \cite{martens2006magnetic}. Experimentally, \citet{laufenberg2006quantitative} have reported on the observation of coupling between rings in patterned arrays, showing that it is possible to realize such systems. More recently, it was demonstrated that it is possible to do basic neuromorphic computations in arrays of connected rings, utilizing the domain walls in these arrays \cite{dawidek2021dynamically}. However, in all of these works, the temperature was not a parameter of interest and concepts like emerging order and phase transitions were not considered. Here, we inspect the temperature dependent magnetic order in this system, and find the formation of clusters, where the amount of bonds between rings can be more than one. Consequently, this system mimics the formation of polymers coupling together a significant number of individual magnetic textures, with the bonding valency set by the multipolarity of a ring.

We used the micromagnetic simulation package \textsc{MuMax3}, which solves the Landau-Lifshitz-Gilbert equation for a grid of cells describing magnetic moments \cite{Vansteenkiste:2014et}. The cell size was $l_x\times l_y \times l_z =$ 2.5 nm $\times$ 2.5 nm $\times$ 4 nm, where $l_z$ is equal to the thickness of the rings. The saturation magnetization and exchange stiffness are $M_{\mathrm{S}}=1\times$10$^6$ A/m and $A_{\mathrm{ex}}=$~1$\times$10$^{-11}$ J/m, and the damping is $\alpha=$ 0.01, effectively describing a material that resembles Permalloy. The rings have an outer and inner radius of $r_o=$ 125 nm and $r_i=$ 75 nm, respectively. The square grid in the simulation provides an effective fourfold anisotropy to the rings due to the corrugated edges, which was partially compensated by a cubic anisotropy set along the $[1,1]$ and $[1,-1]$ directions, resulting in a weak 8-fold anisotropy. The rings were placed on a square 16 $\times$ 16 grid, with a mutual spacing of $d=50$ nm. A finite temperature was taken into account by way of a stochastic field which is proportional to the root of temperature, $\sqrt{T}$, and uncorrelated both in space and time \cite{leliaert2017adaptively}.

\begin{figure}
    \centering
    \includegraphics[width=\columnwidth]{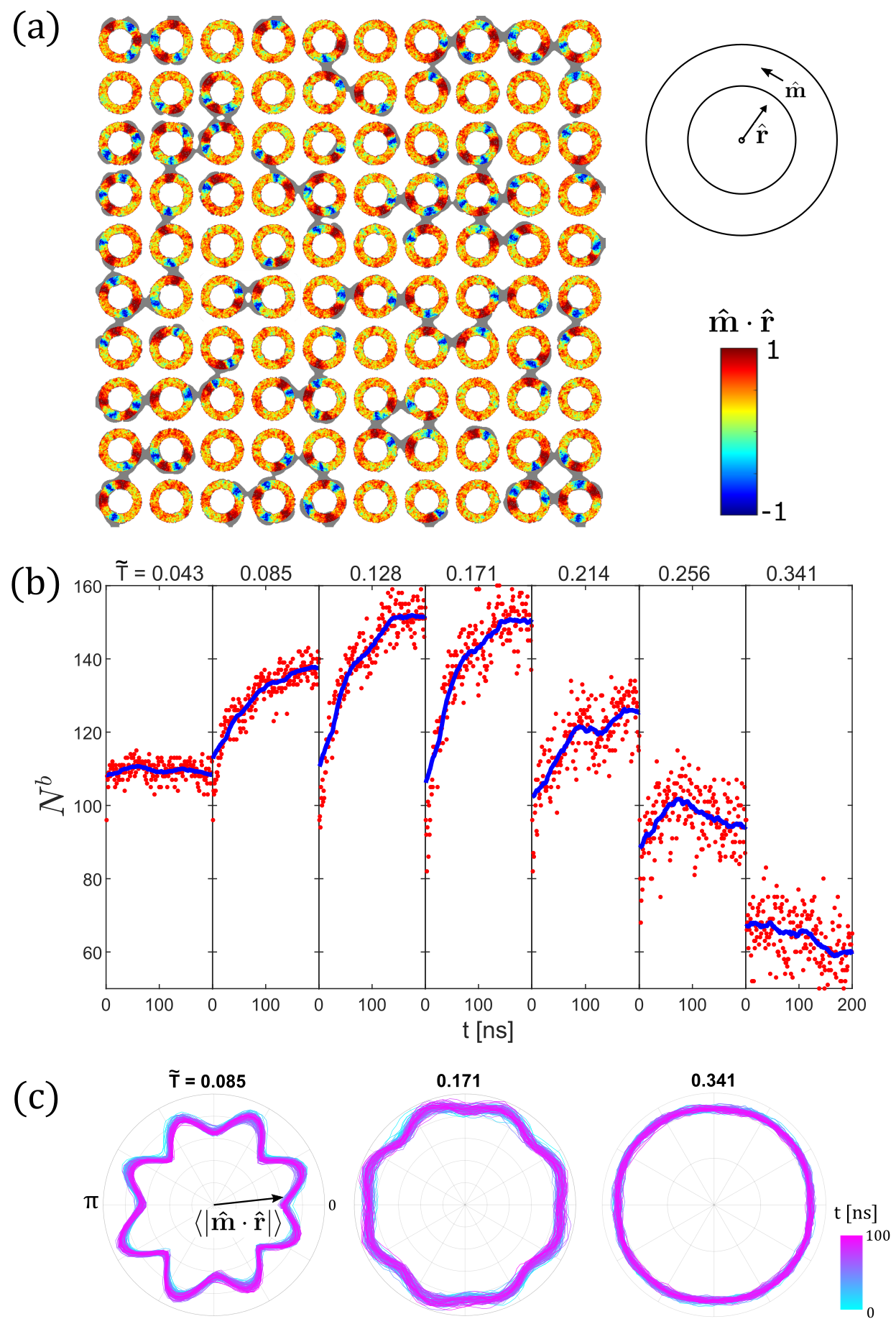}
    \caption{(a) Excerpt from the 16 $\times$ 16 array of rings at $\widetilde{T}=0.171$, after relaxing from a random magnetization. The colormap which is given by the dot product between the normalized radial and magnetization vectors, and represents the magnetostatic charge of the domain wall. The bonding field, $H_{\mathrm{b}}$, is shown in gray. (b) Average cluster size for different reduced temperatures $\widetilde{T} = k_{\mathrm{B}}T/E_{\mathrm{b}}$, starting from a random configuration. The window average of the data is shown in blue, the red dots are the raw data. (c) Time evolution of the average domain wall density, as a function of azimuthal position with respect to the individual magnetic rings (calculated as $\langle|\mathbf{\hat{m}}\cdot\mathbf{\hat{r}}|\rangle$), for three different temperatures.} 
    \label{fig:latticeBonds}
\end{figure}

We set out by considering the magnetization within a single ring. Magnetic rings have a fundamentally different magnetic texture than conventional nanomagnet elements, as the topology is the same as that of a strip with continuous boundary conditions. As such, their coupling to neighbouring elements occurs via the stray field emitted from domain walls, instead of the conventional majority part of the magnetic texture (see Fig. \ref{fig:intro}a). When exposed to temperature, the domain walls are free to move around the magnetic ring, resulting in a key difference compared to magnetic disks, which is the additional freedom of the domain walls to move with respect to each other without a high energy cost. The amount of bonds available for one ring, and thereby the valency, depends on the amount and type of domain walls. A ring with a vortex state has zero bonds, an onion state provides a double valency, and an antivortex state has a quadruple valency. In most of the cases, the sum of the magnetostatic charges is zero, i.e., in the case of the onion state the two magnetostatic charges are $+$ and $-$. This is the case if the winding numbers of the topological defects on the outer edges, as defined by \citet{Tchernyshyov:2005gs}, are positive ($n = +1/2$), and the ones on the inner edges are negative ($n = -1/2$). This configuration is most often the case, since defects with positive topological numbers have the lowest energy on positively curved edges, and vice versa for defects with $n=-1/2$ on negatively curved edges. In some cases, $n=-1/2$ charges can be found on the outer edges of the ring, which results in an uncompensated net magnetostatic charge. In such a case, the state can decay through annihilation of a $n=+1/2$ charge.

The energy landscape of the magnetization in the ring can be mapped out as a function of the total net  magnetization in-plane (its components being $m_x$ and $m_y$), as shown in Fig. \ref{fig:intro}b for the case of two domain walls. In this case, the energy landscape is flat in the azimuthal direction, due to the rotational symmetry of the ring, but curved in the radial direction. The landscape features one deep inner trough and a shallow outer rim. The deep inner trough represents the groundstate, corresponding to the configuration in which the domain walls are close together, due to the attractive interaction between magnetostatic charges, but still remain apart due to the repulsive interaction of the topological charges. This repulsive interaction leads to an upturn in the energy landscape at the smallest net magnetization values. The outer rim in the energy landscape corresponds to a state where the two domain walls are on opposite sides of the ring. 

When magnetic rings are organized in an array and starting from a paramagnetic state, the emergent order is not immediately apparent from just considering the magnetization. As mentioned previously, the rings can couple to one another through the stray fields, which are produced at the domain walls in the magnetic texture in the ring. 
In order to define a bond, we introduce a scalar ``{\it bonding field}", $H_{\mathrm{b}}$, which has binary values, by thresholding the demagnetizing field, $|\mathbf{H}_{\mathrm{dem}}|$:

\begin{equation*}
    \begin{aligned}
        |\mathbf{H}_{\mathrm{dem}}(x,y)|>H_{\mathrm{thresh}} \rightarrow H_{\mathrm{b}}(x,y)= 1 \\
        |\mathbf{H}_{\mathrm{dem}}(x,y)|<H_{\mathrm{thresh}} \rightarrow H_{\mathrm{b}}(x,y)= 0
    \end{aligned}
\end{equation*}

We used $\mu_0H_{\mathrm{thresh}}$ = 0.03 T, and details on the justification of this value and further identification of bonds can be found in the supplementary material. If two rings are connected by $H_{\mathrm{b}}$, this defines a bond. In order to reveal the underlying order, $H_{\mathrm{b}}$ must be included in the visualisation. The appearance of bonds is shown in Fig. \ref{fig:latticeBonds}a, where bond coupling causes clustering into polymers, where the individual rings are considered monomers. As such, the type of order that emerges is of a percolative nature. 

We will now inspect the behaviour of the artificial polymers upon thermal excitation. In this analysis, the order parameter is taken to be the amount of bonds on the lattice, $N_{\mathrm{b}}$. The energy needed to break one bond is $E_{\mathrm{b}} = 0.50$~eV \footnote{This value is calculated by evaluating the demagnetization energy between two rings connected by a bond at 50 nm, and subsequently increasing the distance between the rings in steps of 5 nm until the bond no longer exists.}, and henceforth we will make use of a dimensionless temperature, scaled by this value, $\widetilde{T} = k_{\mathrm{B}}T/E_{\mathrm{b}}$, where $k_{\mathrm{B}}$ is the Boltzmann constant. We have simulated the magnetization for seven different temperatures between $\widetilde{T} = 0.043$ and 0.341, for a duration of 200 ns, in each case starting from the same magnetization state. This initial state has 56 rings with vortex states, 151 rings with two charges 48 rings with four charges, and one ring with six charges. This initial magnetization state is relaxed from a random magnetization in the absence of temperature, and is not the groundstate, but rather a metastable state with $N_{\mathrm{b}}$ $=$ $96$. As time progresses, it can be seen in Fig. \ref{fig:latticeBonds}b that the amount of bonds on the lattice increases for temperatures up to $\widetilde{T}=$ 0.256. In the cases of $\widetilde{T}=$ 0.085 to 0.171, there is a steep increase in $N_{\mathrm{b}}$ during the initial 15-20 ns, after which the rate slows down. We attribute this increase to the annihilation of charges (domain walls) with an inverted winding, thus providing more space for the other charges to move to a bonding position (see supplementary material for the total amount of domain walls over time). An exception is the high temperature case of $\widetilde{T}= 0.341$, where the thermal energy is high enough to annihilate domain walls with regular winding in the timescales of the simulation. Overall, the largest increase in $N_{\mathrm{b}}(t)$ over 200 ns is seen to occur for a temperature of $\widetilde{T}= 0.171$. This behaviour can be rationalized by the fact that, below this temperature, an increased thermal agitation leads to a higher annealing rate, which means that it can be expected that eventually the cases with $\widetilde{T}\le0.171$ relax to low lying metastable minima, with a high $N_{\mathrm{b}}(t)$. At higher temperatures, the entropy begins to prevail, which results in the breaking of bonds and unpairing of charges. As such, we can establish that there exists a ceiling temperature around $\widetilde{T}^* = 0.171$, analogous to the ceiling temperature in realistic polymer systems, at which the rate of polymerization equals that of depolymerization. This behaviour signifies a soft transition, involving the unpairing of charges, thereby bearing resemblance to a Berezinskii-Kosterlitz-Thouless phase transition \cite{1971JETP_Berezinski, kosterlitz1972long}. At the critical temperature, the system optimizes the mobility of the domain walls to find bonding positions. A detailed investigation of this transition is due in future work.

The difference in behaviour of the system when the temperature is varied is ultimately reflected in the positions of the domain walls, whose motion is governed by the underlying energy landscape. The domain wall positions are tracked, and a histogram was made as a function of their position in polar angle, for all timesteps, shown in Fig. \ref{fig:latticeBonds}c. At low temperatures, the domain wall positions strongly reflect the 8-fold anisotropy of the individual rings. When the temperature is increased, the system has enough energy to overcome this anisotropy, and the domain wall position is mostly dominated by the lattice symmetry, as seen from the four-fold anisotropy of the histogram. When the temperature is increased further to $\widetilde{T} = 0.341$, we observe that the domain walls explore the full phase space, unimpeded by local or global anisotropies. As such, the system becomes fully ergodic at the highest temperature.

\begin{figure}
    \centering
    \includegraphics[width=\columnwidth]{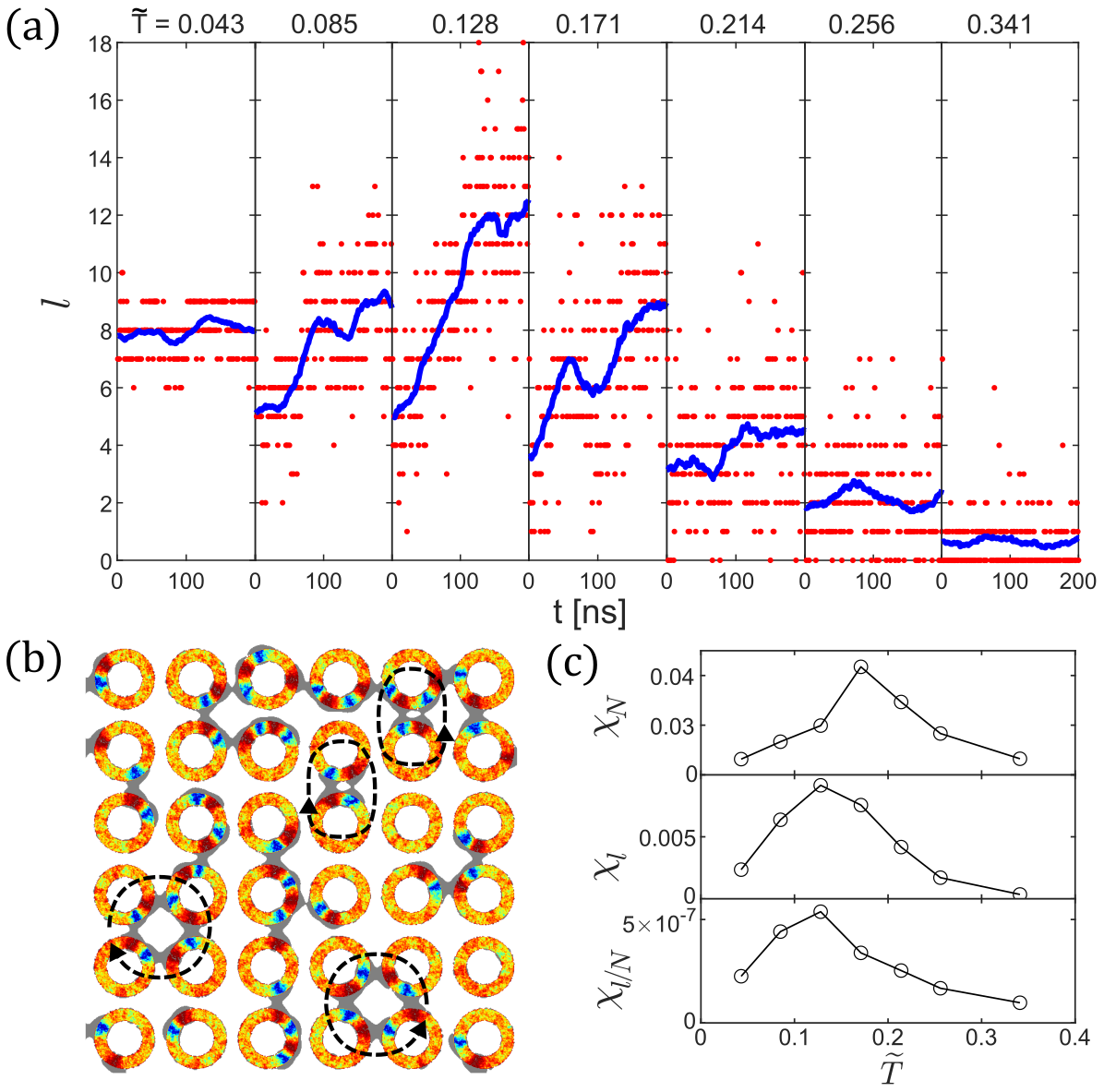}
    \caption{(a) Amount of loop-clusters for different temperatures, starting from a random configuration. The blue line indicates the windowed average, the red dots are the raw data (b) Example of a loop clusters spanning 2 and 4 rings. (c) The susceptibility is shown for different temperatures for the amount of bonds (upper panel), the amount of loops (middle panel), and for the amount of loops normalized by the amount of bonds (lower panel).}
    \label{fig:thermloop}
\end{figure}

We observe that the tendency of the ring system to compensate the magnetic charges leads to the formation of emergent topological structures on the next length scale, namely loops, as can be seen in Fig. \ref{fig:thermloop}b. The typical size of these loops is two and four rings. Loops play an important role in conventional artificial spin ice systems, where they realize a topological model system, in which trivial and non-trivial loops can be distinguished \cite{nisoli2020topological,schanilec2022approaching}. The (possibly degenerate) groundstate of the ring lattice must feature only loops, in order for all charges to be compensated. As such, we expect an increased amount of loops at lower temperatures. This tendency to form loops is enhanced by the intra-ring interactions, which favour small angles between domain walls, causing the emerging polymers to bend. This can be contrasted to a lattice of disks, which favours straight lines due to the contribution of the exchange interaction in the interior \cite{Ewerlin:2013dv, vedmedenko2005multipolar, sloetjes2017tailoring, digernes2020direct}. Moreover, rings with four domain walls serve as pinch points in the polymers, also promoting the formation of loops. In the following, we shall investigate the thermal dynamics of these loops.

We start by considering the average polymer length, given by $\langle L\rangle$. This parameter is closely related to $N_{\mathrm{b}}$. However, if one is to relate the two quantities mathematically, the amount of loops, $l$, must also be taken into account:

\begin{equation}
    \langle L\rangle = \frac{N}{N-N_{\mathrm{b}}+l}
\end{equation}

\noindent
where $N$ is the total amount of rings. As such, once $\langle L\rangle$ and $N_{\mathrm{b}}$ are found, $l$ can be determined via this relationship. The amount of loops over time is shown for different temperatures in Fig. \ref{fig:thermloop}a. For lower temperatures, initially the amount of loops drops drastically (for $t=0$), before increasing again. The initial relative drop in loops ($\Delta l/l(t=0)$) is much larger than that for the amount of bonds ($\Delta N_{\mathrm{b}}/N_{\mathrm{b}}(t=0)$). This can perhaps be attributed to a different relaxation time associated with loop formation versus bond formation. The total amount of loops fluctuates between 0 and 18, where the maximum is seen for $\widetilde{T}=0.128$, in contrast to the number of bonds, which stabilizes at $\widetilde{T}=0.171$. This contrast suggests a different critical temperature associated with bond formation versus that of loop formation. We investigate this possibility by inspecting the susceptibility (see Fig. \ref{fig:thermloop}c), which is calculated as $\chi_x(T) = (1/k_{\mathrm{B}}T) \sigma_{x}^2$, where $\sigma_{x} = \sqrt{\langle x^2\rangle - \langle x\rangle^2}$ is the standard deviation of $x$, with $x=N_{\mathrm{b}}$ or $l$ \cite{newman1999monte}. A peak in the susceptibility is associated with a maximum in the fluctuations, and could indicate the occurrence of a phase transition. We observe peaks in the susceptibilities for both $\chi_{N_{\mathrm{b}}}$ and $\chi_l$. However, the susceptibility of the bonds peaks at $\widetilde{T}=\widetilde{T}^*=$ 0.171, while the susceptibility peak for the loops occurs at $\widetilde{T}$ = 0.128. The susceptibility for the loops was also calculated when normalized by the amount of bonds, $\chi_{l/N_{\mathrm{b}}}$, and it was found that the peak remained localized at $\widetilde{T}$ = 0.128. The difference in critical temperature for these two entities is surprising, as the loops consist of bonds, and one would therefore expect the same critical temperature. However, there is one crucial difference: whereas bond formation only depends on inter-ring interactions, the formation of loops depends on both inter- and intra-ring interactions, thus setting a different energy scale. Additionally, the entropy associated with $N_{\mathrm{b}}$ and $l$ should be considered. The configurational entropy associated with placing a bond on a lattice is smaller than the entropy associated with placing structures on the lattice that consist of multiple bonds, such as loops, thus reducing the critical temperature for $l$ with respect to $N_{\mathrm{b}}$.

In conclusion, we have investigated a novel artificial spin system, in which the topology of the individual building blocks is altered with respect to conventional magnetic metamaterials. We have found that bonding between elements occurs via stray fields emitted and absorbed by domain walls, giving rise to a magnetic order that resembles polymerization. Moreover, the formation of bonds is associated with a thermodynamic transition. On the next length scale, we observe an additional transition associated with the formation of a topological structure known as a loop, with a definite handedness. The critical temperatures associated with the transitions of these two entities differ, which we attribute to different energy scales and configurational entropy. 
\\

\noindent
The data that support the reported findings are available upon reasonable request.

\section{Acknowledgements}
\noindent
We wish to thank Prof. Bj\"orgvin Hj\"orvarsson for fruitful discussions. S.D.S. and V.K. acknowledge support from the Swedish Research Council (Project No. 2019-03581). M.P.G. and V.K. also acknowledge support from the Carl Trygger Foundation (Project No. CTS21:1219).
\\

\noindent
The authors have no conflicts of interest to disclose

%

\pagebreak
\onecolumngrid
\newpage
\begin{center}
\textbf{\large Supplemental Material: Polymerization in magnetic metamaterials}
\end{center}
\setcounter{equation}{0}
\setcounter{figure}{0}
\setcounter{table}{0}
\setcounter{page}{1}
\makeatletter
\renewcommand{\theequation}{\bf S\arabic{equation}}
\renewcommand{\figurename}{Supplementary FIG.}
\renewcommand{\thefigure}{{\bf S\arabic{figure}}}
\renewcommand{\bibnumfmt}[1]{[S#1]}
\renewcommand{\citenumfont}[1]{S#1}
\renewcommand{\thepage}{S-\arabic{page}}

\section{Justification of the bonding field}

\begin{figure}[t]
    \centering
    \includegraphics[width=0.9\textwidth]{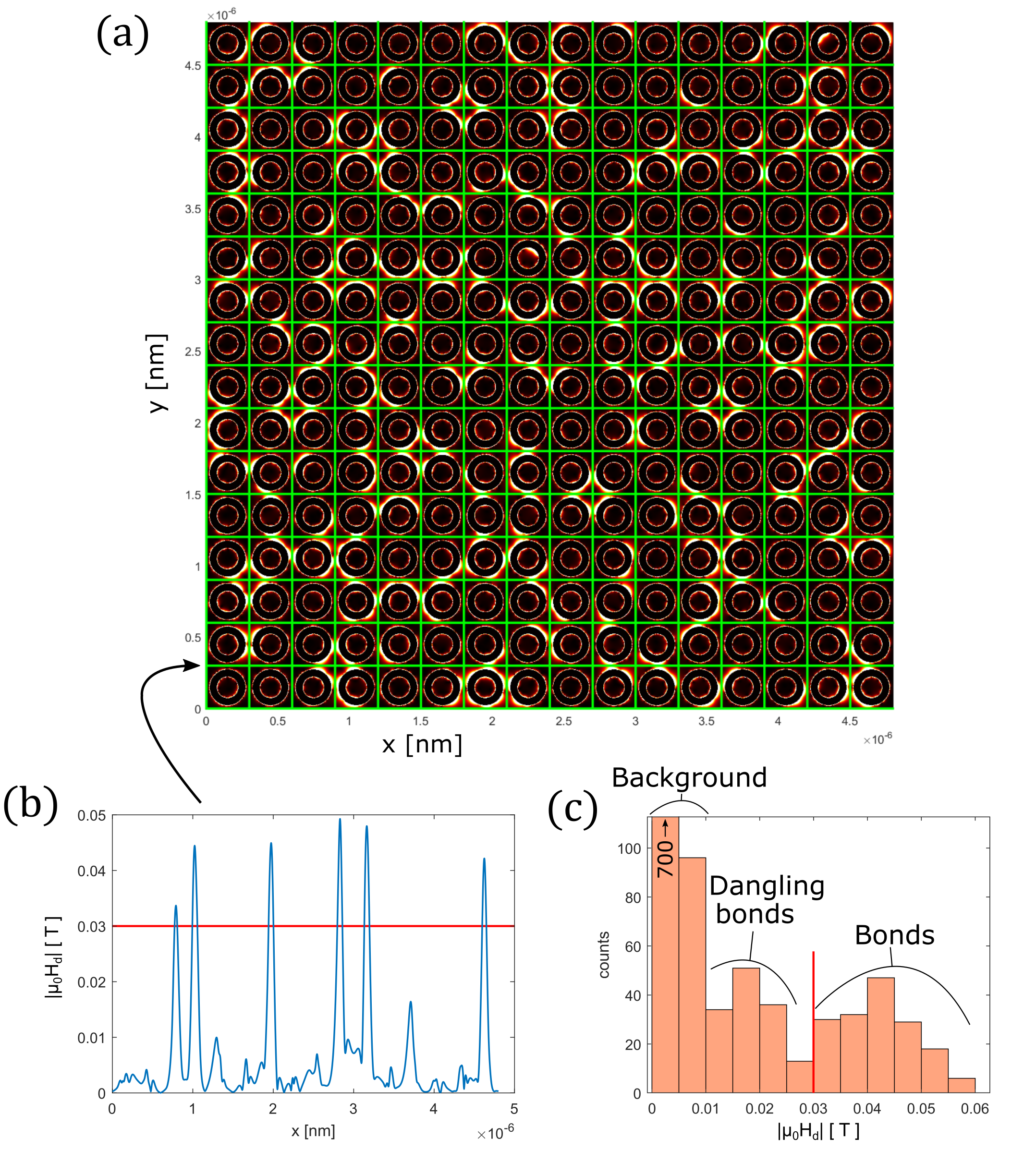}
    \caption{(a) Intensity plot of the demagnetizing field, $|\mu_0H_\mathrm{d}|$ outside the rings. The green lines indicate the positions at which the cross-sections were evaluated. (b) Example of a cross-section of $|\mu_0H_\mathrm{d}|$ along the line indicated with a black arrow. The red line indicates the threshold for the classification as bond. (c) Trimodal distribution of the peak heights in (b), for all lines.
    }
    \label{fig:size}
\end{figure}

The choice for a threshold of 30 mT was made based on the histogram of peak heights in the demagnetizing field. To construct this histogram, we evaluated the demagnetizing field along horizontal and vertical lines that divided the spacing between the rings up in two equal parts, see Fig. \ref{fig:size}a. The idea behind this is that a bond has a high demagnetizing field in between the nanomagnets. Figure \ref{fig:size}b shows an example of the demagnetizing field along one of the lines.
By plotting the histogram of peak heights in the demagnetizing field along all the lines, we observe that there is a trimodal distribution of peak heights, i.e. we recognize three characteristic groups. The high values of dipolar field correspond to domain wall that face each other, i.e. bonds. The second peak under 30 mT corresponds to individual domain walls without a partner in the adjacent ring, i.e. unsatisfied dangling bonds. Finally, the peak at low dipolar field corresponds to the background expected for a long range interaction. 

\newpage 

\section{Identification of bonds}

\begin{figure}[t]
    \centering
    \includegraphics[width=0.9\textwidth]{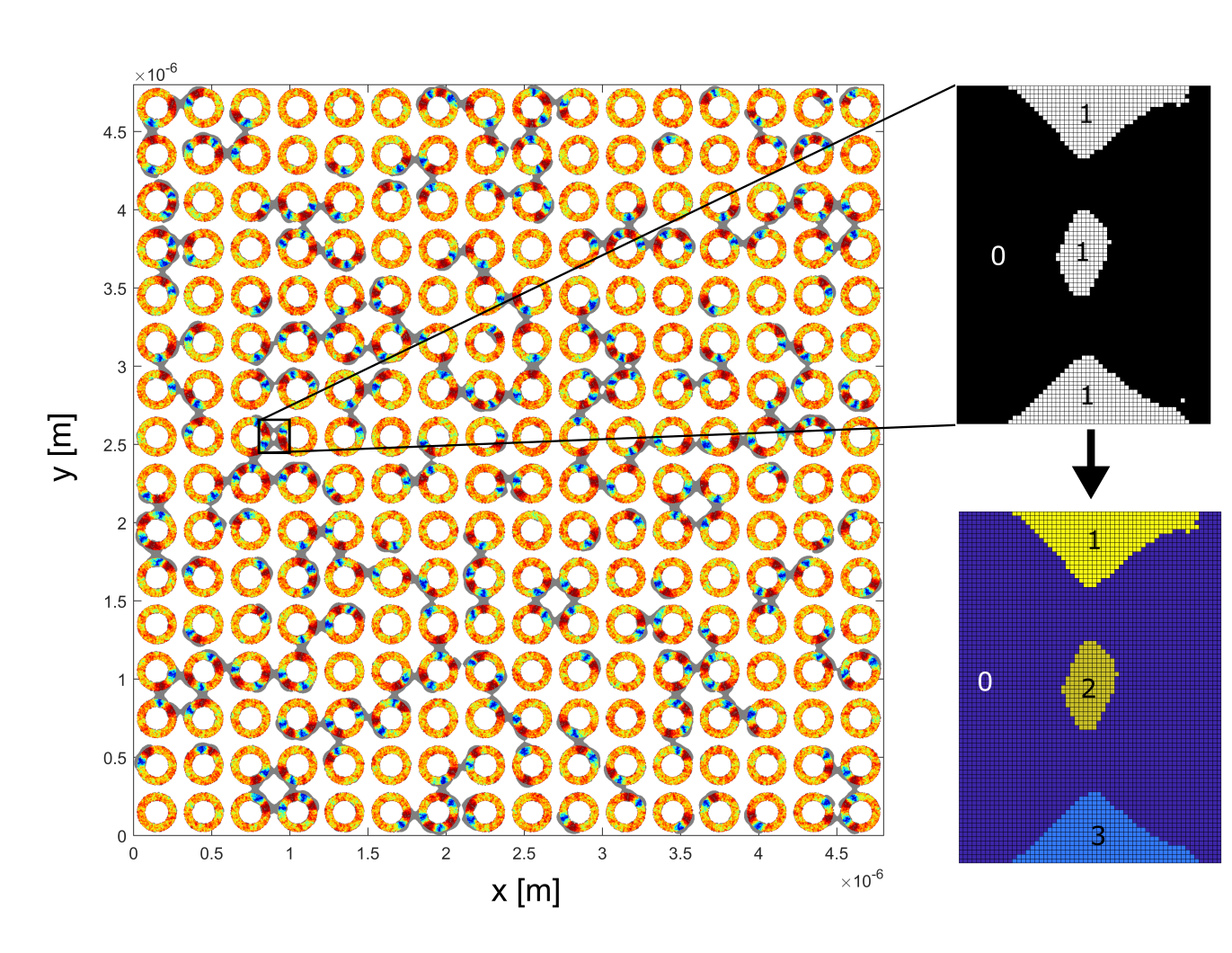}
    \caption{Identification of bonds. Left: the system of rings and bonds at $\widetilde{T}=$ 0.128, right: zoom in of the region with the bonds, after the labelling using MATLAB.
    }
    \label{fig:bondID}
\end{figure}

The identification of bonds was done by cutting out rectangular regions of size 150 $\times$ 200 nm in between the rings, and inspecting the bonding field. From each rectangle, a binary image was made, in which the regions of the ring, and regions with $H_b=1$, are mapped to a value 1, and the rest to 0, see Fig. \ref{fig:bondID}. Then, using the MATLAB function `\texttt{bwlabel}', the different connected regions where labeled. If there are only 2 different regions (0 and 1), this indicates the absence of a bond. In the case that there were 3 regions (0, 1 and 2), a single bond was identified, and in the case of four different regions (see Fig. \ref{fig:bondID}), a double bond was present. In order to distinguish from artifacts, only regions with 10 or more connected pixels were taken into account. This procedure was carried out in two steps; one for the horizontal bonds, and one for the vertical bonds.
\newpage 

\section{Domain wall number evolution}

\begin{figure}[t]
    \centering
    \includegraphics[width=0.8\textwidth]{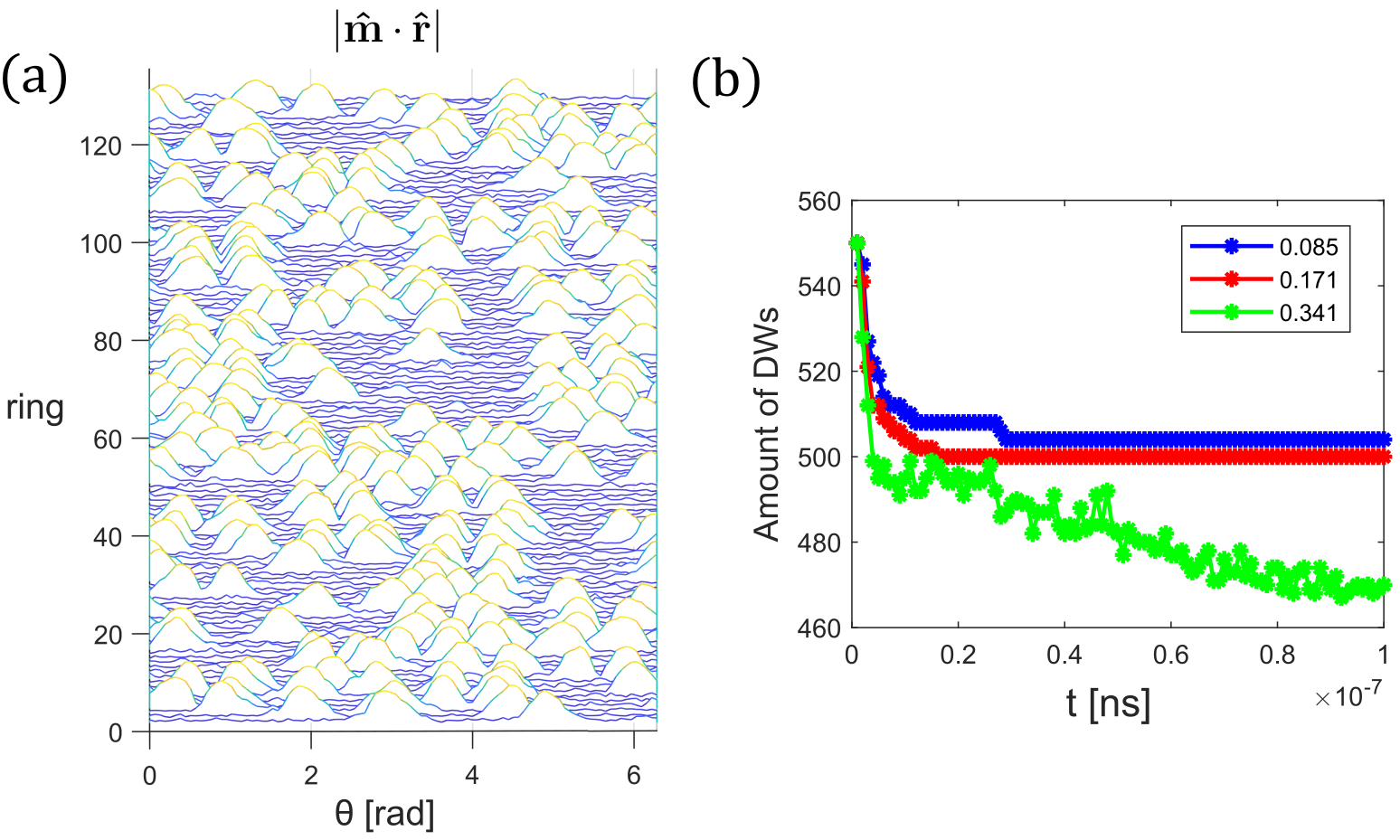}
    \caption{(a) Waterfall plot of $|\mathbf{\hat{m}}\cdot\mathbf{\hat{r}}|$ against the angle, $\theta$ in each ring, for 128 rings, the peaks indicate the presence of a domain wall. The temperature in this case is $\widetilde{T}=0.085$. (b) Total amount of domain walls over time, for three different temperatures.
    }
    \label{fig:DWnum}
\end{figure}

The domain wall population was quantified and plotted against time, see Fig. \ref{fig:DWnum}b. This was done by plotting for each ring $|\mathbf{\hat{m}}\cdot\mathbf{\hat{r}}|$ against angle, see Fig \ref{fig:DWnum}a, and using the MATLAB `\texttt{findpeaks}' function to identify the peaks, as each peak indicates a domain wall.

\end{document}